\documentclass{svproc}
\usepackage{url}

\usepackage{amsmath,graphicx}

\usepackage{enumitem}
\usepackage{multirow} 
\usepackage{booktabs}

\begin{document}
\mainmatter              
\title{Improving Diagnostic Accuracy of Pigmented Skin Lesions With CNNs: A Case Study with DermaMNIST Dataset}
\titlerunning{Improving Diagnostic Accuracy of Pigmented Skin Lesions With CNNs: an Application on the DermaMNIST Dataset}  
%
\author{Nerma Kadric \and Amila Akagic \and Medina Kapo}
\authorrunning{Nerma Kadric et al.} 
%
\tocauthor{Nerma Kadric, Amila Akagic, Medina Kapo}
\institute{Faculty of Electrical Engineering, University of Sarajevo,\\
Computer Science and Informatics\\
71000 Sarajevo, Bosnia and Herzegovina\\
\email{\{nkadric1, aakagic, mkapo2\}@etf.unsa.ba}}

\maketitle              

\begin{abstract}
Pigmented skin lesions represent localized areas of increased melanin and can indicate serious conditions like melanoma, a major contributor to skin cancer mortality. The MedMNIST v2 dataset, inspired by MNIST, was recently introduced to advance research in biomedical imaging and includes DermaMNIST, a dataset for classifying pigmented lesions based on the HAM10000 dataset. This study assesses ResNet-50 and EfficientNetV2L models for multi-class classification using DermaMNIST, employing transfer learning and various layer configurations. One configuration achieves results that match or surpass existing methods. This study suggests that convolutional neural networks (CNNs) can drive progress in biomedical image analysis, significantly enhancing diagnostic accuracy.
\end{abstract}

\section{Introduction}
\label{sec:intro}
Pigmented skin lesions represent a prevalent dermatological finding, frequently identified by patients through self-examination. These lesions manifest as localized areas of increased melanin deposition, resulting in spots or patches that appear darker than the adjacent skin. Although there are different types of skin diseases, special focus is given to tumors, with melanoma being the most common form of tumor, responsible for a high percentage of deaths worldwide.

Skin changes result from various factors including infections, allergies, autoimmune diseases, injuries, genetics, hormonal variations, and sun exposure. In the U.S. and globally, skin cancer rates and mortality are rising~\cite{ring2021dermatoscopy}. Key types of skin cancer include melanoma, basal cell carcinoma (BCC), and squamous cell carcinoma (SCC). Pre-cancerous conditions like actinic keratosis and Bowen's disease also require attention~\cite{malvehy2006handbook},~\cite{mackie1996skin},~\cite{nikolouzakis2020current}. Actinic keratoses~\cite{marghoob2005atlas},~\cite{agnew2015fast},~\cite{qin2020gan}, are precancerous conditions resulting from long-term sun exposure, while intraepithelial carcinomas and basal cell carcinomas represent different forms of skin cancer, with basal cell carcinomas often occurring in sun-exposed areas of the skin. Benign lesions such as dermatofibroma are usually not dangerous, but can cause aesthetic discomfort or irritation. On the other hand, melanoma is a serious form of skin cancer that can appear in any part of the body, and requires prompt diagnosis and aggressive treatment. Melanocytic nevi are common lesions, benign moles that can vary in size, shape, and color, while vascular lesions include various conditions such as hemangiomas, telangiectasias, and vascular malformations.

Recently, a new dataset collection MedMNIST v2 was published~\cite{medmnistv2}.  The objective of this dataset is to support a wide range of research and educational initiatives in the field of biomedical imaging. The authors drew their motivation to create this dataset from the MNIST dataset, a foundational resource widely used in the early stages of computer vision research. MedMNIST v2 comprises 12 datasets for 2D imaging and 6 datasets for 3D imaging. Among these, the DermaMNIST dataset is derived from a comprehensive collection of multi-source dermatoscopic images of common pigmented skin lesions, encompassing various lesion types such as melanoma and basal cell carcinoma, based on the HAM10000 dataset~\cite{ham1}. Same as HAM10000, DermaMNIST consists of 10,015 dermatoscopic images categorized as 7 different diseases. 

Dermatoscopy \cite{malvehy2006handbook}, ~\cite{marghoob2005atlas}, as an advanced technique for analyzing skin diseases, focuses on detecting pigmented changes on the skin's surface. Automated classification systems accelerate the workflow and identification of different types of skin diseases. In recent years, Deep Learning algorithms are used for analysis of biomedical images. Convolutional neural networks are the most prominent type of network that achieves the best results. CNNs made a significant breakthrough when GoogleNet used this technology for cancer detection with an accuracy of 89\%, while human pathologists achieved only 70\% accuracy~\cite{huang2018applications}.
 

In this paper, we examine the effectiveness of two state-of-the-art models, ResNet-50 and EfficientNetV2L, in feature extraction from the DermaMNIST dataset. Using transfer learning, we extract features from the images and apply various layer configurations for multi-class classification. One of these configurations achieves results that are higher or comparable to those of existing state-of-the-art methods.

\begin{figure}
  \centering
\includegraphics[width=0.9\textwidth,height=0.35\textwidth]{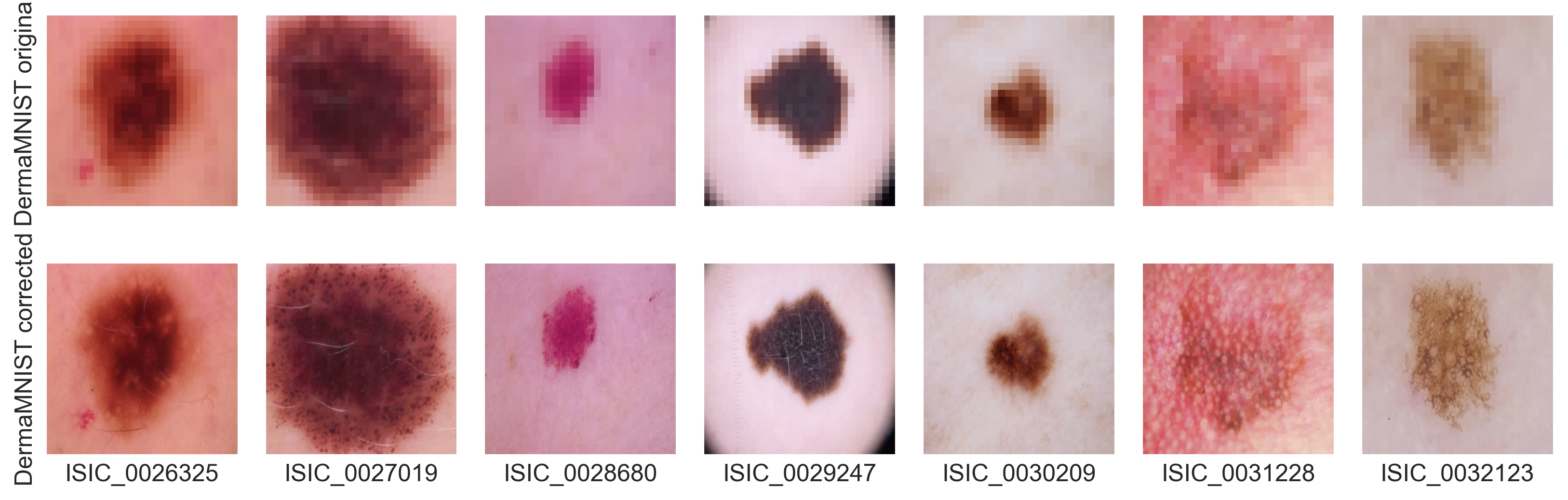}
  \caption{The first row represents the original DermaMNIST images, while the second row represents the DermaMNIST-C images.}
\label{fig:slikederma}
\end{figure}

\section{Datasets}
\label{sec:typestyle}
DermaMNIST~\cite{yang2021medmnist}, ~\cite{dermamnist1}, ~\cite{dermamnist2} is a resized version of HAM10000, with images reduced from 600x450 to 28x28 pixels. Images are classified in seven classes: 1) Actinic Keratoses and Intraepithelial Carcinom, 2) Basal Cell Carcinoma, 3) Benign Keratosis-like Lesions, 4) Dermatofibroma, 5) Melanoma, 6) Melanocytic Nevi, and 7) Vascular Lesions. Images are split into training, validation, and test sets in the range of 70:10:20\%. 

The dataset has faced criticism for poor organization and class imbalance. In response, authors Abhishek et al. proposed an improved version, 
called\\ DermaMNIST-C~\cite{abhishek2024investigating}, which features enhanced image resolution (224x224) and improved organization. This version includes 8,208 training images, 1,232 test images, and 575 validation images, addressing issues of duplication and resolution to enhance classification outcomes. Fig.~\ref{fig:slikederma} illustrates the differences between the DermaMNIST and DermaMNIST-C datasets.

\section{Methods}
\label{sec:pagestyle}
In this paper, we explore the capabilities of leading state-of-the-art models, including EfficientNetV2L and ResNet-50, for multi-class classification of images in the DermaMNIST and DermaMNIST-C datasets. Our experiments primarily employ transfer learning; however, we also investigate the impact of adding various layers atop the base model to enhance classification performance.

Two distinct deep learning architectures, EfficientNetV2L and ResNet-50, both pre-trained on ImageNet, are employed for feature extraction. The base model weights are frozen, and additional layers are integrated, including a flatten layer, batch normalization, dense layers with varying units (64 or 128), dropout, and a final dense layer with 7 softmax units for multi-class classification.

The modified models are trained using categorical cross-entropy loss and the Adam optimizer (learning rate = 1e-4). To ensure optimal training, EarlyStopping and ModelCheckpoint callbacks are applied, with model performance assessed via accuracy/loss curves and a confusion matrix. The architectures leverage activation functions, including ReLU and Softmax, with hyperparameter optimization playing a crucial role in refining model performance. The primary objective is to achieve accurate classification of unseen data.


\section{Metrics}
\label{sec:metr}
A range of metrics is employed to assess classifier performance and the effectiveness of data representation within a model. Selecting appropriate metrics is crucial in the evaluation of deep learning models, as certain metrics may present a misleading performance overview if misapplied. Thus, we use Precision, Accuracy, Recall and AUC. Some metrics are only included to compare results with other state-of-the-art methods. 

Precision measures the ratio of true positive predictions to the total number of positive predictions. It is particularly useful in situations where minimizing false positives is crucial. Mathematically, precision is given as Eq. (\ref{eq:prec})~\cite{buza2022unsupervised}.

\begin{equation}
    \text{Precision}=\frac{TP}{TP + FP}
    \label{eq:prec}
\end{equation}

Accuracy represents the ratio of correct predictions to the total number of predictions. For binary classification, it is given as Eq. (\ref{eq:acc})~\cite{buza2022unsupervised}.

\begin{equation}
\text{Accuracy (ACC)} = \frac{TP + TN}{TP + FP + TN + FN}
    \label{eq:acc}
\end{equation}

Recall quantifies the proportion of actual positive samples correctly identified. This metric is important in contexts where minimizing false negatives is critical, such as in medical diagnostics. Recall is defined as Eq. (\ref{eq:recc})~\cite{buza2022unsupervised}.
\begin{equation}
    \text{Recall}=\frac{TP}{TP + FN}
        \label{eq:recc}
\end{equation} 

AUC (Area Under the Curve) is used in ROC (Receiver Operating Characteristic) analysis to rank classifier performance. It represents the area under the ROC curve, with values closer to 1 indicating better performance. AUC is calculated by math Eq. (\ref{eq:auc}), where $S_p$ shows sum of ranks all positives, $n_p$ and $n_n$ number of positive and negative examples~\cite{hossin2015review}.

\begin{equation}
\text{AUC} = \frac{S_p - n_p(n_p + 1)/2}{n_p \cdot n_n}
    \label{eq:auc}
\end{equation}

\section{Results}\label{sec:results}
Two experiments are created to assess the capabilities of models. In the first experiment, we perform classification on DermaMNIST, and in the second on the DermaMNIST-C dataset. 

\subsection{Experiment 1: Classification on DermaMNIST dataset}
In this experiment, we implement an identical classification configuration atop the base ResNet-50 and EfficientNetV2L models. This configuration consists of two blocks, each containing two consecutive Conv2D layers with 32 and 64 filters, respectively, followed by MaxPooling2D and Dropout layers. These are succeeded by two additional Conv2D layers with 128 filters, followed by Dropout, a Flatten layer, Batch Normalization, a Dense layer, another Dropout and Batch Normalization layer, and a final Dense layer with 7 neurons for classification.

Table~\ref{tab:tab1} presents the classification results for the initial approach using enhanced ResNet50 and EfficientNetV2L models on DermaMNIST images. The EfficientNetV2L model demonstrated improvements over ResNet50, with better ACC, precision, and AUC metrics.

\begin{table}
    \centering
    \begin{tabular}{lccccc}
        \toprule
        & Loss & ACC & Precision  & AUC & Recall\\
        \midrule
        Res50\_e  &\textbf{ 1.2829}          & 0.6663 &  0.7152    & 0.9178 & \textbf{0.6299} \\
        Eff\_e  &  1.2933            & \textbf{0.7017}   & \textbf{0.8977}  & \textbf{0.9334} &0.5556\\
        \bottomrule
    \end{tabular}
    \vspace{3pt}
        \caption{Results - DermaMNIST 28x28x3}

    \label{tab:tab1} 
\end{table}

\subsection{Experiment 2: Classification on DermaMNIST-C dataset}
In this experiment, we evaluate five different models. The first, termed Simple Model (SM), employs the same configuration as the prior models from the experiment 1 but omits the base models, resulting in a model trained from scratch on the DermaMNIST-C dataset. The second model, Effv1\_e, replicates the configuration of the first experiment’s model but is also trained on the DermaMNIST-C dataset. The third model, Effv2\_e, simplifies the classification structure by including only a Flatten layer after the base model, followed by Batch Normalization, a Dense layer with 128 neurons, another Batch Normalization layer, and a final Dense layer with 7 neurons for classification. In the fourth configuration, Effv3\_e, we retain the architecture of Effv2\_e but reduce the Dense layer neurons to 64. Finally, in Effv4\_e, we use the same configuration as Effv3\_e but incorporate class weights to address the dataset's class imbalance.

Table~\ref{tab:tab2} shows the results for the simple model and various EfficientNetV2L (Effv1\_e, Effv2\_e, Effv3\_e, Effv4\_e) models. EfficientNetV2L versions provided significant improvements, with v3 achieving the highest ACC and recall.

\begin{table}
    \centering
    \begin{tabular}{lccccc}
    \toprule
         & Loss & ACC & Precision  & AUC & Recall\\
        \midrule
        SM     & 1.6109     & 0.8117  & 0.8480   & \textbf{0.9775} & 0.7922 \\
        Effv1\_e  &  0.9206  & 0.8271   & \textbf{0.9324 }        & 0.9711 & 0.7394\\
        Effv2\_e  & 0.9873  &0.8401     & 0.8798         & 0.9768 & 0.8076\\
        Effv3\_e & \textbf{0.8808}  &\textbf{0.8490}  &0.8935  & 0.9768 & \textbf{0.8239}\\
        Effv4\_e  & 1.8290  &0.8247              & 0.8740             & 0.9666  & 0.7768\\
        \bottomrule
    \end{tabular}
    \vspace{3pt}
    \caption{Results - DermaMNIST-C 224x224x3}
    \label{tab:tab2} 
\end{table}

\section{Discussion}\label{sec:discussion}
\begin{figure}
  \centering
  \begin{minipage}{0.46\textwidth}
    \includegraphics[width=\textwidth]{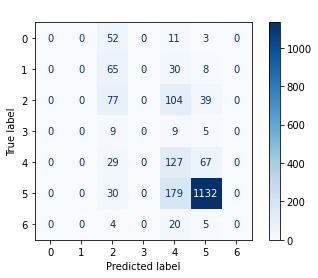}
        \caption{Confusion matrix of ResNet model for the Experiment 1.}
        \label{fig:resnet}
  \end{minipage}
  \hfill
  \begin{minipage}{0.45\textwidth}
    \includegraphics[width=\textwidth]{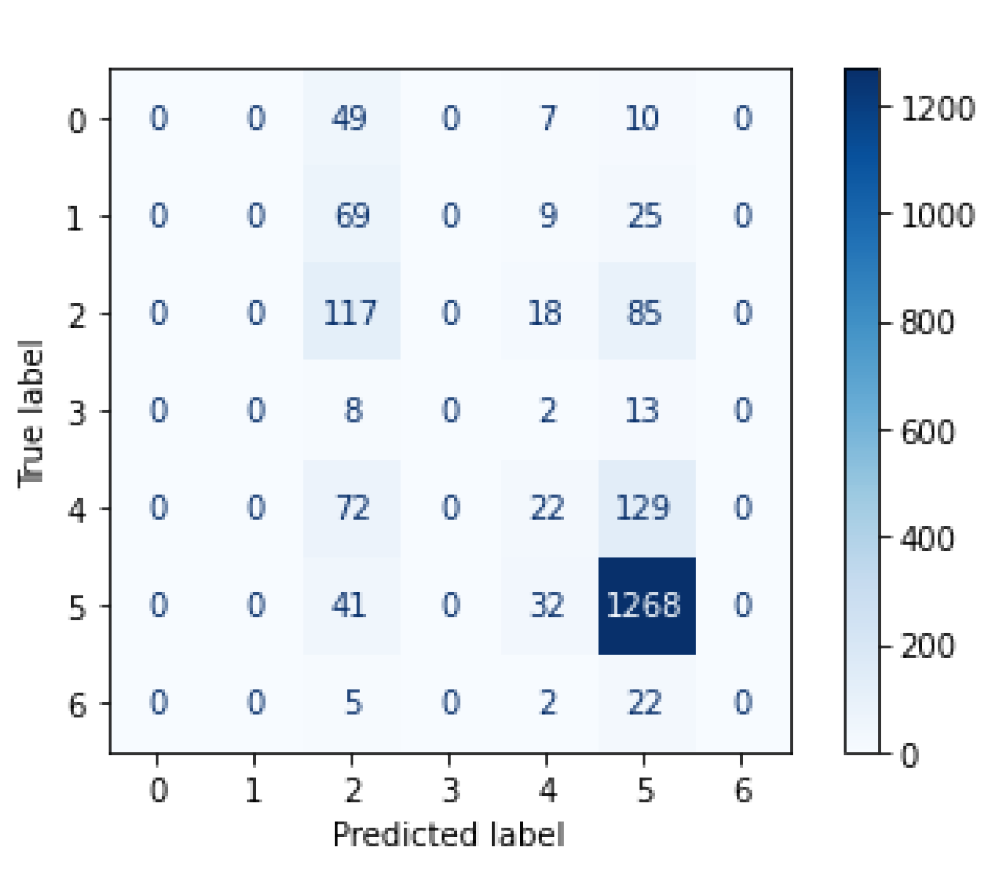}
   \caption{Confusion matrix of EfficientNet model for the Experiment 1.}
        \label{fig:effnet_exp1}
  \end{minipage}
\end{figure}

The first experiment using the ResNet-50 and EfficientNetV2L models on the DermaMNIST dataset revealed issues with overfitting, as seen in poor accuracy on the test subset and high number of false positives in the confusion matrix (Fig.~\ref{fig:resnet} and Fig.~\ref{fig:effnet_exp1}). This indicated limitations presented in the dataset, including class imbalance and poor organization. The model's performance remained suboptimal, with several classes poorly predicted. Overall, these results underscored that the DermaMNIST dataset was inadequate for reliable classification. 

\begin{figure}
  \centering
    \includegraphics[width=0.45\textwidth]{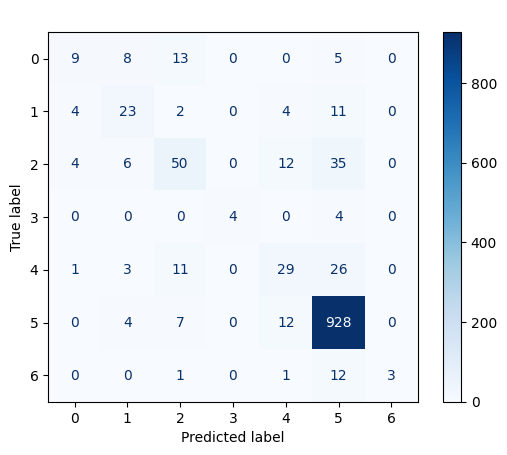}
        \caption{Confusion matrix of EfficientNetV2L v3 model for the Experiment 2.}
        \label{fig:effnet_exp2}
\end{figure}

In the second experiment, the DermaMNIST-C dataset was employed, with one of the resulting confusion matrices displayed in Fig.~\ref{fig:effnet_exp2}. Analysis of this confusion matrix and the associated results reveals an improvement in recall with this configuration. Importantly, the imbalance in the number of images per class continues to substantially impact the model's performance characteristics.

In Table~\ref{tab:tab3}, a comparison with different state-of-the-art methods is made. Results demonstrate that EfficientNetV2L model achieved impressive accuracy of 0.8490 on DermaMNIST-C dataset. This improvement is attributed to innovative modeling approach, with the contribution of feature extraction capabilities of EfficientNetV2L as well. Comparison with other methods such as ResNet50 model from~\cite{abhishek2024investigating} shows a slightly higher accuracy of 0.851, however, as is seen in Table 1 and Table 2, higher accuracy does not reflect to higher values in other metrics such as Precision, AUC and Recall. Thus, EfficientNetV2L model also showed competitive and exceptional performance in classifying complex dermatological data.

\section{Conclusion}\label{sec:conclusion}
This study explored application of advanced deep learning techniques for classifying dermatological images. Given high prevalence of skin conditions including melanoma aim was to enhance classification accuracy and efficiency using advanced machine learning models. The research underscores importance of utilizing sophisticated deep learning techniques in medical image analysis. It lays groundwork for future development and improvement of diagnostic tools. These tools can significantly benefit patient care. 

Due to imbalanced nature of DermaMNIST dataset, future work could involve expanding dataset with generated images for each class. This may further enhance classification results. Additionally integrating such classification models into medical applications could aid in detecting malignant and benign skin changes. This would contribute to better healthcare outcomes.

\begin{table}
    \centering
    \begin{tabular}{lccc}
        \toprule
        & Dataset & Method & ACC \\
        \midrule
         \cite{yang2021medmnist} & DermaMNIST  & GAutoML Vision & 0.766 \\
\cite{abhishek2024investigating} & \textit{DermaMNIST-C} & \textit{ResNet-50} &  \textit{\textbf{0.851}}\\
        \cite{win2020hybrid} & DermaMNIST & GAutoML Vision & 0.768 \\
        \cite{app12052634} & DermaMNIST  & AOC-Caps & 0.786 \\
         \cite{mercaldo2024extreme} & DermaMNIST & ELM & 0.674 \\
         \cite{yang2024skin} & DermaMNIST & - & 0.826 \\
         \cite{xu2022medrdf} & DermaMNIST & ResNet-18 & 0.741 \\
         \cite{ahmed2022failure} & DermaMNIST & ResNet-18 & 0.7367 \\
         \cite{jiang2022deeply} & DermaMNIST & LSANet+SimAM & 0.7940 \\
                \multirow{2}{*}{\textbf{This}}  & DermaMNIST & EfficientNetV2L & 0.7017 \\
                \textbf{work} & \textbf{DermaMNIST-C} & \textbf{EfficientNetV2L} & \textbf{0.8490}\\
        \bottomrule
    \end{tabular}
    \vspace{3pt}
        \caption{Comparison with state-of-the-art methods.}
    \label{tab:tab3} 
\end{table}

\bibliographystyle{IEEEbib}
\bibliography{refs}

\end{document}